\documentstyle[preprint,aps,graphicx,floats]{revtex}

\begin{document}

\preprint{$
\begin{array}{l}
\mbox{CERN-TH/2003-186}\\[-3mm]
\mbox{UB-HET-03-06}\\[-3mm]
\mbox{DESY-03-103}\\[-3.mm]
{\rm October}~2003 \\ [3mm]
\end{array}
$}

\title{Probing the Higgs self-coupling at hadron colliders \\ using rare
  decays}

\author{U.~Baur\footnote{e-mail: baur@ubhex.physics.buffalo.edu, but
    only when electric power is available\\[-12.mm]}}
\address{Department of Physics,
State University of New York, Buffalo, NY 14260, USA\\[-2.mm]}
\author{T.~Plehn\footnote{e-mail: tilman.plehn@cern.ch, but only when AFS
is operational\\[-12.mm]}}
\address{CERN Theory Group, CH-1211 Geneva 23, Switzerland\\[-2.mm] }
\author{D.~Rainwater\footnote{e-mail: david.rainwater@desy.de, but only
when AFS is operational}}
\address{DESY Theorie, Notkestrasse 85, D-22603 Hamburg,
Germany\\[-3.mm]}

\maketitle

\begin{abstract}
\baselineskip13.pt 
We investigate Higgs boson pair production at hadron colliders for
Higgs boson masses $m_H\leq 140$~GeV and rare decay of one of the two
Higgs bosons.  While in the Standard Model the number of events is
quite low at the LHC, a first, albeit not very precise, measurement of
the Higgs self-coupling is possible in the $gg\to HH\to
b\bar{b}\gamma\gamma$ channel.  A luminosity-upgraded LHC could
improve this measurement considerably.  A 200~TeV VLHC could make a
measurement of the Higgs self-coupling competitive with a
next-generation linear collider.  In the MSSM we find a significant
region with observable Higgs pair production in the small $\tan\beta$
regime, where resonant production of two light Higgs bosons might be
the only hint at the LHC of an MSSM Higgs sector.
\end{abstract}

\newpage


\tightenlines

\section{Introduction}
\label{sec:sec1}

The CERN Large Hadron Collider (LHC) is scheduled to begin operation
in 2007, beginning a new era wherein the mechanism of electroweak
symmetry breaking and fermion mass generation will be revealed and
studied in great detail.  Although alternative mechanisms exist in
theory, this is generally believed to be a light Higgs boson with mass
$m_H<219$~GeV~\cite{lepewwg}.  More specifically, we expect a
fundamental scalar sector which undergoes spontaneous symmetry
breaking as the result of a potential which acquires a nonzero vacuum
expectation value.  The LHC will easily find a light Standard Model
(SM) Higgs boson with very moderate luminosity~\cite{wbf_ww,wbf_exp}.
Moreover, the LHC will have significant capability to determine many
of its properties~\cite{atlas_tdr,cms_tdr}, such as its fermionic and
bosonic decay modes and couplings~\cite{wbf_ll,Hcoup,Yb,Yt}, including
invisible decays~\cite{wbf_inv} and possibly even rare decays to
second generation fermions~\cite{Hmumu}.~\footnote{An $e^+e^-$ linear
  collider with a center of mass energy of 350~GeV or more can
  significantly improve these preliminary measurements, in some cases
  by an order of magnitude in precision, if an integrated luminosity
  of 500~fb$^{-1}$ can be achieved~\cite{LC}.}

Starting from the requirement that the Higgs boson has to restore
unitarity of weak boson scattering at high energies in the
SM~\cite{unit}, perhaps the most important measurement after a Higgs
boson discovery is of the Higgs potential itself, which requires
measurement of the trilinear and quartic Higgs boson self-couplings.
Only multiple Higgs
boson production can probe these directly~\cite{higgs_self,LC_HH1}.
\medskip

Recent literature is replete with self-coupling measurement studies.
There are numerous quantitative sensitivity limit analyses of Higgs
boson pair production in $e^+e^-$ collisions ranging from 500~GeV to
3~TeV center of mass energies~\cite{LC_HH1,LC_HH2,LC_HH3,LC_HH4}.  For
example, one neural net-based study concludes that a 500~GeV linear
collider with an integrated luminosity of 1~ab$^{-1}$~\cite{LC_HH4}
could measure the trilinear Higgs coupling $\lambda$ for $m_H=120$~GeV,
where $H\to b\bar{b}$ decays 
dominate, at the $20\%$ level.  However, none of these analyses
addressed the case of $m_H>140$~GeV, where the Higgs boson mostly
decays into $W$ bosons.  Studies exploring the potential of
the LHC, a luminosity-upgraded LHC (SLHC) with roughly
ten times the amount of data expected in the first run, and a
Very Large Hadron Collider (VLHC), have come only very
recently~\cite{SLHC,BPR,blondel,Baur:2003gp}.  These studies
investigated Higgs pair production via gluon fusion with subsequent
decay to same-sign dileptons and three leptons via $W$ bosons, and
cover the broader range $115<m_H<200$~GeV.  They established that
future hadron machines can probe the Higgs potential for $m_H\gtrsim
150$~GeV.  At the LHC, an integrated luminosity of 300~fb$^{-1}$
provides for exclusion of vanishing $\lambda$ at the $95\%$ confidence
level or better over the entire range $150<m_H<200$~GeV.  A VLHC
would provide for precision measurement over much of this mass range,
similar to or better than the limits achievable at a 3~TeV $e^+e^-$
collider with 5~ab$^{-1}$~\cite{LC_HH3}.  However,
we previously concluded that hadron colliders could
not probe the mass region $m_H<140$~GeV sufficiently well to be
meaningful~\cite{Baur:2003gp}.
\medskip

We reexamine that conclusion in this paper, utilizing rare decay modes
in Higgs
boson pair production for $m_H<140$~GeV at future hadron colliders.
We first review the definition of the
Higgs boson self-couplings and briefly discuss SM and non-SM
predictions for these parameters in Sec.~\ref{sec:theory}.  An
overview of the rare Higgs decay modes in the SM (predominantly
$b\bar{b}\gamma\gamma$ final states) and our analyses of these
channels appears in Sec.~\ref{sec:lhc}.  We consider the LHC, SLHC and
a VLHC, which we assume to be a $pp$ collider operating at 200~TeV
with a luminosity of ${\cal L}=2\times 10^{34}~{\rm cm^{-2}\,
  s^{-1}}$~\cite{vlhc}.  In Sec.~\ref{sec:mssm} we establish
the prospects of observing a pair of minimal supersymmetric Standard
Model (MSSM) Higgs bosons in the $b\bar{b}\gamma\gamma$ and
$b\bar{b}\mu^+\mu^-$ decay channels.  We present our conclusions in
Sec.~\ref{sec:conc}.

\section{Higgs boson self-couplings}
\label{sec:theory}

The trilinear and quartic Higgs boson couplings $\lambda$ and
$\tilde\lambda$ are defined through the potential
\begin{equation}
\label{eq:Hpot}
V(\eta_H) \, = \,
{1\over 2}\,m_H^2\,\eta_H^2\,+\,\lambda\, v\,\eta_H^3\,+\,{1\over 4}\,
\tilde\lambda\,\eta_H^4 ,
\end{equation}
where $\eta_H$ is the physical Higgs field, $v=(\sqrt{2}G_F)^{-1/2}$
is the vacuum expectation value, and $G_F$ is the Fermi constant.  In
the SM the self couplings are
\begin{equation}
\label{eq:lamsm}
\tilde\lambda=\lambda=\lambda_{SM}={m_H^2\over 2v^2}\,.
\end{equation}
Regarding the SM as an effective theory, the Higgs boson
self-couplings $\lambda$ and $\tilde\lambda$ are {\it per se} free
parameters, and $S$-matrix unitarity constrains $\tilde\lambda$ to
$\tilde\lambda\leq 8\pi/3$~\cite{unit}.  Since future collider
experiments likely cannot probe $\tilde\lambda$, we concentrate on the
trilinear coupling $\lambda$ in the following.  The quartic Higgs
coupling does not affect the Higgs pair production processes we
consider.

\medskip

In the SM, radiative corrections decrease $\lambda$ by $4-11\%$ for
$120< m_H<200$~GeV~\cite{yuan}.  Larger deviations are
possible in scenarios beyond the SM. For example, in two Higgs doublet
models where the lightest Higgs boson is forced to have SM like
couplings to vector bosons, quantum corrections may increase the
trilinear Higgs boson coupling by up to $100\%$~\cite{yuan}.  In the
MSSM, loop corrections modify
the self-coupling of the lightest Higgs boson in the decoupling limit,
which has SM-like 
couplings, by up to $8\%$ for light stop squarks~\cite{hollik}.
Anomalous Higgs boson self-couplings also appear in various other
scenarios beyond the SM, such as models with a composite Higgs
boson~\cite{georgi}, or in Little Higgs models~\cite{lhiggs}.  In
many cases, the anomalous Higgs boson self-couplings can be
parameterized in terms of higher dimensional operators which are
induced by integrating out heavy degrees of freedom.  A systematic
analysis of Higgs boson self-couplings in a higher dimensional
operator approach can be found in Ref.~\cite{tao}.

\section{Analysis}
\label{sec:lhc}

At LHC energies, inclusive Higgs boson pair production is dominated by
gluon fusion~\cite{lhc_hh}.  Other processes, such as weak boson
fusion, $qq\to qqHH$~\cite{wbf}, associated production with heavy
gauge bosons, $q\bar{q}\to WHH, ZHH$~\cite{assoc}, or
associated production with top quark pairs, $gg,\,q\bar{q}\to
t\bar{t}HH$~\cite{SLHC}, yield cross sections which are factors of
10--30 smaller than that for $gg\to HH$~\cite{lhc_hh}.  Since $HH$
production at the LHC is generally rate limited, we consider only the
gluon fusion process.

\begin{figure}[t!]
\begin{center}
\includegraphics[width=12cm,height=9cm]{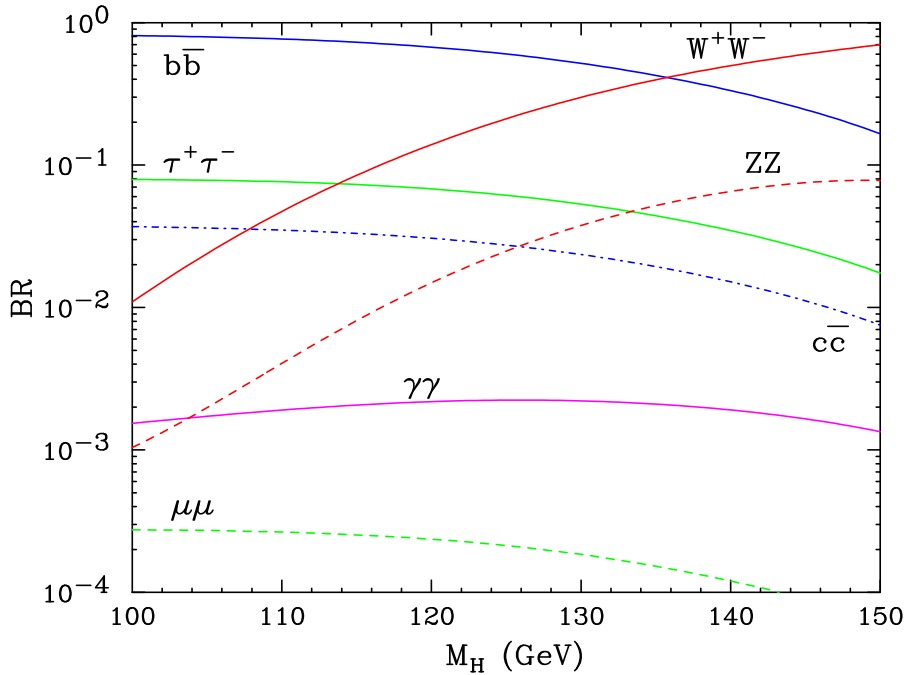}
\vspace*{2mm}
\caption[]{\label{fig:BRs} SM Higgs branching ratios relevant to our 
  analysis of $HH$ production.  For $W^+W^-$ and $ZZ$, one of the
  gauge bosons is off-shell.}  \vspace{-7mm}
\end{center}
\end{figure}

Because the total $gg\to HH$ cross section at both the LHC and VLHC is
quite small, at most one
Higgs boson undergoing rare decay will allow for a reasonable number of
events to work with.  We therefore consider only final states
containing one $b$-quark pair, which is the dominant SM Higgs boson
decay mode for $m_H<135$~GeV, as shown in Fig.~\ref{fig:BRs}.  Our
previous study demonstrated that at both LHC and
VLHC, $4b$ and $b\bar{b}\tau^+\tau^-$ final states are overwhelmed 
by backgrounds~\cite{Baur:2003gp}.  While the backgrounds are more
moderate for the $\tau$-channel, the 
observable part of this decay mode unfortunately has multiple
additional small branching ratios, and the detectors have rather low
efficiency to identify the $\tau$-leptons.  As charm quarks are even
more difficult to tag than $b$-quarks, and the QCD backgrounds become
much larger due to similarly less fake-tag rejection, we can
immediately discount any colored final states for the rare decay.
Weak boson pairs certainly qualify as rare decays in this
mass region, but cannot be used: the $b\bar{b}W^*W$ and $b\bar{b}
Z^*Z\to b\bar{b}\ell^+\ell^-\bar\nu\nu$ final states suffer from a huge
QCD top pair background.  Similarly for $pp\to HH\to b\bar{b}Z^*Z$ with
one or more hadronically decaying $Z$ bosons, and $b\bar{b}Z\gamma\to
b\bar{b}jj\gamma$, QCD processes with the same final states are likely
to overwhelm the signal (here, $W^*$ and $Z^*$ denote off-shell $W$
and $Z$ bosons).  The $b\bar{b}Z^*Z\to b\bar{b}+4$~leptons and
$b\bar{b}Z\gamma\to\ell^+\ell^-\gamma$ channels suffer from too low a
rate, due to the small $Z\to\ell^+\ell^-$ branching ratio.  This
leaves only the diphoton $b\bar{b}\gamma\gamma$ and dimuon
$b\bar{b}\mu^+\mu^-$ decay combinations.\smallskip

For all our calculations we assume an integrated luminosity of
600~fb$^{-1}$ for the LHC, and 6000~fb$^{-1}$\cite{SLHC} for the SLHC.
For the VLHC, we consider both 600~fb$^{-1}$ and
1200~fb$^{-1}$~\cite{vlhc}.  We choose
$\alpha_s(M_Z)=0.1185$~\cite{long}, calculate signal and background
cross sections using CTEQ5L~\cite{Lai:1999wy} parton distribution
functions, and our scale choice for all background processes is
$\mu_F=\mu_R=\sqrt{\hat s}$.  We include minimal detector effects by
Gaussian smearing of the parton momenta according to ATLAS
expectations~\cite{atlas_tdr}, and take into account energy loss in
the $b$-jets via a parameterized function.  We assume a $b$-tagging
efficiency of $\epsilon_b=50\%$ for all hadron colliders.  In
addition, we include an efficiency of $79\%$~\cite{sasha} for
capturing the $H\to b\bar{b}$ decay of the signal in its 40~GeV mass
bin.  We calculate all background processes using {\sc
madgraph}~\cite{madgraph} except where otherwise noted, and retain a
finite $b(c)$-quark mass of 4.6(1.7)~GeV where relevant.  Other
detector efficiencies are given in the subsections relevant to
the respective channels.

\subsection{The $b\bar{b}\gamma\gamma$ decay channel}
\label{sec:gamgam}

We perform the signal calculation, $gg\to HH\to b\bar{b}\gamma\gamma$,
as in Refs.~\cite{BPR,Baur:2003gp}, including the effects of
next-to-leading order (NLO) QCD corrections via a multiplicative
factor $K=1.65(1.35)$ at LHC(VLHC) energies~\cite{hh_nlo}, using
factorization and renormalization scales choices of $m_H$.  There is
little scale variation left at NLO. We use exact matrix elements to
incorporate the $H\to b\bar{b}$ and $H\to\gamma\gamma$
decays.

The basic kinematic acceptance cuts for events at the (S)LHC and VLHC
are:
\begin{eqnarray}
\label{eq:cuts1}
\nonumber &
p_T(b) > 45~{\rm GeV} \; , \qquad
|\eta(b)| < 2.5 \; , \qquad
\Delta R(b,b) > 0.4 \; , \\
\nonumber &
m_H-20~{\rm GeV} \, < \, m_{b\bar{b}} \, < \, m_H+20~{\rm GeV} \; , \\
\nonumber &
p_T(\gamma) > 20~{\rm GeV} \; , \qquad
|\eta(\gamma)| < 2.5 \; , \qquad
\Delta R(\gamma,\gamma) > 0.4 \; , \\
\nonumber &
m_H-2.3~{\rm GeV} \, < \, m_{\gamma\gamma} \, < \, m_H+2.3~{\rm GeV} \;
, \\ &
\Delta R(\gamma,b) > 0.4 \; ,
\end{eqnarray}
which are motivated first by requirements that the events can pass the
ATLAS and CMS triggers with high efficiency~\cite{atlas_tdr,cms_tdr},
and that the $b$-quark and photon pairs reconstruct to windows around
the known Higgs boson mass, adjusted for an expected capture
efficiency of $79\%$ each~\cite{sasha}.  We take the identification
efficiency for each photon to be $80\%$ at all machines
considered~\cite{sasha}.\medskip

As in the $4W$ signal case~\cite{BPR}, we will later try to determine
the Higgs boson self-coupling from the shape of the invariant mass of
the final state.  For that reason we do not apply any cuts which make
use of the fact that the signal involves two heavy massive particles
produced in a fairly narrow range of the $b\bar{b}\gamma\gamma$
invariant mass.
The only irreducible background processes are
QCD $b\bar{b}\gamma\gamma$, $H(\to\gamma\gamma)b\bar{b}$ and $H(\to
b\bar{b})\gamma\gamma$ production.
However, there are multiple QCD reducible backgrounds
resulting from jets faking either $b$-jets or photons:
\begin{itemize}
\item[$\cdot$] $c\bar{c}\gamma\gamma$ - one or two fake $b$ jets;
\item[$\cdot$] $b\bar{b}j\gamma$ - one fake photon;
\item[$\cdot$] $c\bar{c}j\gamma$ - one or two fake $b$-jets, one fake photon;
\item[$\cdot$] $jj\gamma\gamma$ - one or two fake $b$-jets;
\item[$\cdot$] $b\bar{b}jj$ - two fake photons;
\item[$\cdot$] $c\bar{c}jj$ - one or two fake $b$-jets, two fake photons;
\item[$\cdot$] $jjj\gamma$ - one or two fake $b$-jets, one fake photon;
\item[$\cdot$] $jjjj$ - one or two fake $b$-jets, two fake photons;
\item[$\cdot$] $Hjj$ - one or two fake $b$-jets, or two fake photons;
\item[$\cdot$] $Hj\gamma$ - one fake photon.
\end{itemize}
Misidentified charm quarks must be considered separately from
non-heavy flavor jets because of the grossly different rejection
factors.  Table~\ref{rejfac} summarizes the expected rejection factors
for charm and light jets to be misidentified as $b$-jets and photons,
as well as the expected photon and muon identification efficiencies.
The probability to misidentify a light jet as a $b$-jet is
significantly higher at the SLHC due to the high-luminosity
environment~\cite{SLHC}.  The value quoted in Table~\ref{rejfac} for
$P_{j\to b}$ at the LHC is likely to be conservative; recent
studies~\cite{hawk} using three dimensional $b$-tagging have found a
light jet rejection factor about a factor two better.  Expectations
for the probability to misidentify a light jet as a photon at the LHC
vary considerably~\cite{atlas_tdr,cms_tdr,schwem,abdullin}, so we
perform two analyses, one conservative and the other optimistic, to
cover this range.  Since their design luminosities are similar, it is
reasonable to assume that the rejection factors for light quarks and
charm quarks, and the jet-photon misidentification probabilities, are
similar for the LHC and the VLHC.  Studies of how the high luminosity
environment of the SLHC affects $P_{c\to b}$ and $P_{j\to\gamma}$ have
not yet been performed.  In lieu of better estimates we therefore use
the same values as for the LHC and VLHC.  It should be noted that the
rejection factors listed in Table~\ref{rejfac} depend on the
transverse momentum of the charm quark, $p_T(c)$, or jet, $p_T(j)$.
The values listed in the Table correspond to the rejection factor in
the $p_T$ range which provides the largest contribution to the cross
section.
\begin{table}
\caption[]{\label{rejfac}
Expected photon and muon identification efficiencies, and
misidentification probabilities for charm quarks and light jets as
$b$-quarks~\cite{atlas_tdr,cms_tdr,SLHC,hawk} and
photons~\cite{atlas_tdr,cms_tdr,schwem,abdullin}, at various hadron
colliders.}
\vspace{2mm}
\begin{tabular}{ccccccc}
\phantom{i} & $\epsilon_\gamma$ & $\epsilon_\mu$
& $P_{c\to b}$ & $P_{j\to b}$
& $P^{hi}_{j\to\gamma}$ & $P^{lo}_{j\to\gamma}$ \\[1mm]
\hline\\[-3.5mm]
LHC  & $80\%$ & $90\%$ & 1/13 & 1/140 & 1/1600 & 1/2500 \\
SLHC & $80\%$ & $90\%$ & 1/13 & 1/23  & 1/1600 & 1/2500 \\
VLHC & $80\%$ & $90\%$ & 1/13 & 1/140 & 1/1600 & 1/2500 \\
\end{tabular}
\end{table}

Except for the $b\bar{b}j\gamma$ and $b\bar{b}jj$ backgrounds, all
reducible backgrounds depend on whether one requires one or both
$b$-quarks to be tagged.  Requiring only one tagged $b$-quark results
in a signal cross section which is a factor $(2/\epsilon_b-1)=3$
larger than the one with both $b$-quarks tagged.  This larger
signal rate comes at the expense of a significantly increased
reducible background.  As we shall demonstrate, the small $gg\to HH\to
b\bar{b}\gamma\gamma$ cross section forces us to require a single
$b$-tag at the LHC in order to have an observable signal.  At the
SLHC, on the other hand, the much higher probability to misidentify a
light jet as a $b$-jet translates into an increase of the background
which more than compensates the signal gain from using only a single
$b$-tag.  In the following we therefore require a double $b$-tag at
the SLHC.  For the VLHC we consider both single and double
$b$-tagging.\smallskip

For a single $b$-tag strategy, there is an additional combinatorical
background when extra jets are present in the event.  To estimate this
background, one needs to interface the $gg\to HH$ matrix elements with
an event generator.  Insight may
also be gained from performing a calculation of $HHj$ production,
which presently does not exist.  Since we calculate the signal cross
section with cuts only at lowest order, we do not include the
combinatorical background in our background estimate.

\begin{figure}[t!]
\begin{center}
\vspace{-5mm}
\includegraphics[width=8cm,height=5.8cm]{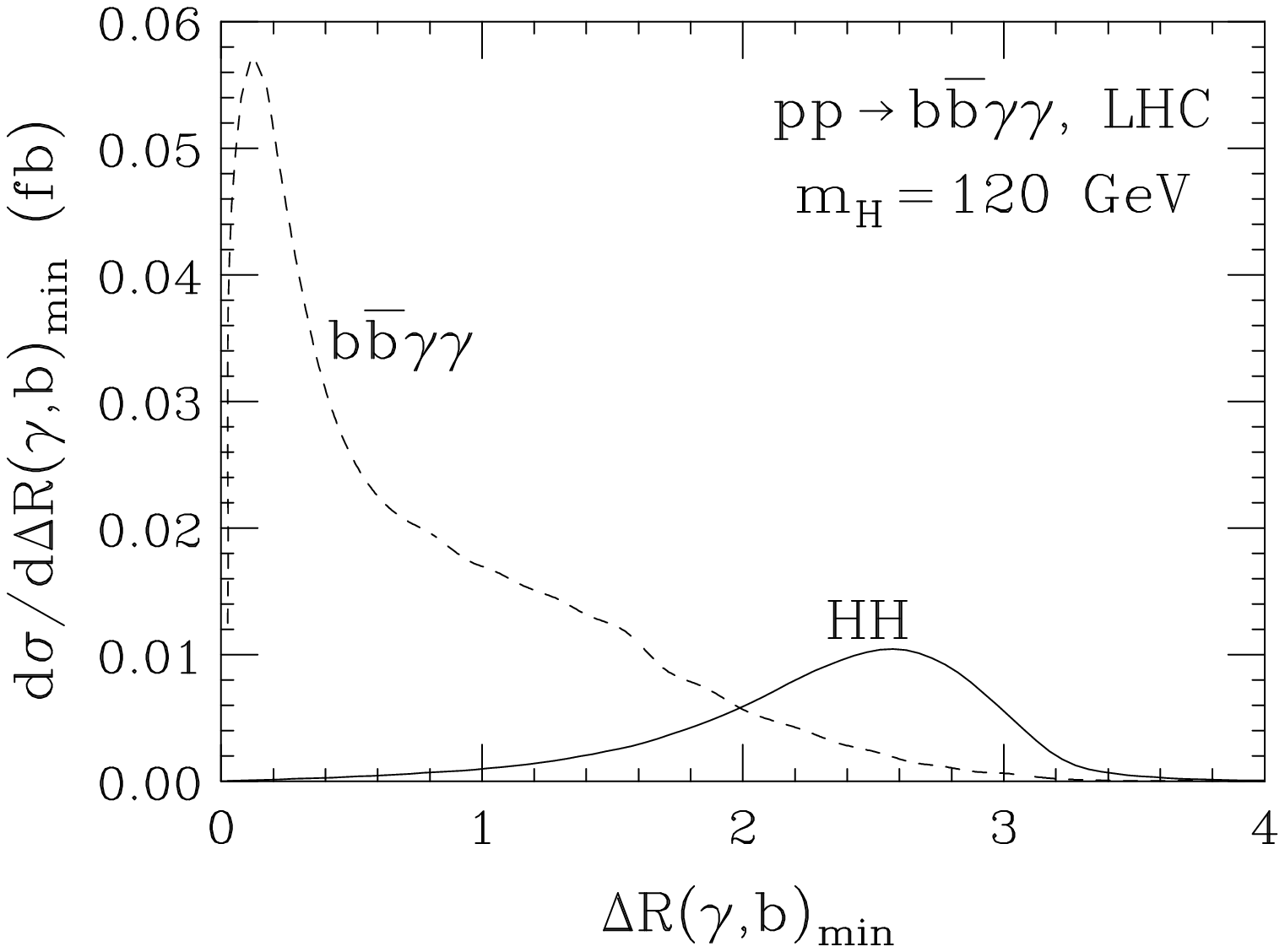}
\includegraphics[width=8cm,height=5.8cm]{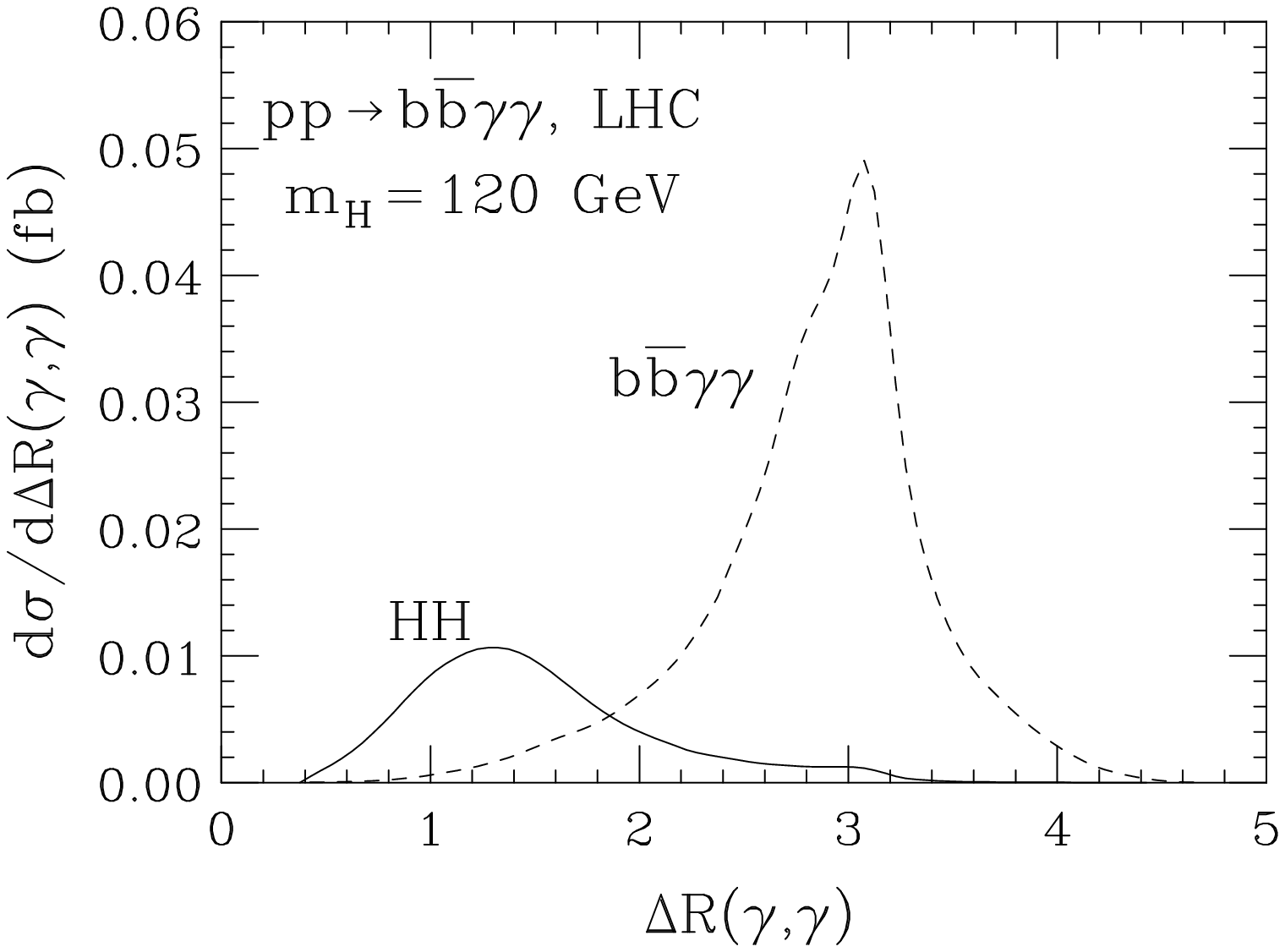}
\vspace*{2mm}
\caption[]{\label{fig:dR}
  Distributions of the minimum lego plot (pseudorapidity -- transverse
  plane) separation between (a) $b$-jets and photons, and (b) the
  photons, for a SM signal of $m_H=120$~GeV and the QCD
  $b\bar{b}\gamma\gamma$ background; using the cuts of
  Eq.~(\protect\ref{eq:cuts1}) but no minimum $b-\gamma$ separation.
  We include the NLO K-factor for the signal and a factor 1.3 for the
  QCD background.}
\vspace{-7mm}
\end{center}
\end{figure}

At the level of cuts in Eq.~(\ref{eq:cuts1}), we observe two angular
correlations which differ strongly between signal and background.  The
minimum separation between $b$-jets and photons is typically much
smaller for the QCD backgrounds as compared to the signal.  The shape
of the signal distribution reflects the fact that the $b\bar{b}$ and
$\gamma\gamma$ pairs originate from decays of heavy scalar particles
which recoil against each other in the transverse plane.  The peak in the
background $\Delta R(\gamma,b)_{min}$ distribution at small
values is clearly due to the collinear enhancement from photon
radiation off a $b$-quark.  The minimum separation between the
photons, on the other hand, is smaller for the signal.  We show the
minimum photon--$b$ and the photon--photon separation distributions in
Fig.~\ref{fig:dR}, for the $HH$ signal and the $b\bar{b}\gamma\gamma$
background at the LHC; all other background processes exhibit
distributions qualitatively similar to those for QCD
$b\bar{b}\gamma\gamma$ production.
Based on these observations, we impose two additional angular cuts on
the final state, which reduce the backgrounds by about an order of
magnitude, but affect the signal at only the $15-20\%$ level for
$m_H=120$~GeV, and closer to $30\%$ for $m_H=140$~GeV:
\begin{equation}
\label{eq:cuts2}
\Delta R(\gamma,b) > 1.0 \; , \qquad \Delta R(\gamma,\gamma) < 2.0 \; .
\end{equation}
Looking at Fig.~\ref{fig:dR}, these to not appear to be the optimum
values.  However, the cuts are correlated, and we chose these values
to roughly optimize $S/B$ while retaining a significant fraction of
the signal.\smallskip

Tables~\ref{tab:xsec.l} and~\ref{tab:xsec.v} display the signal and
QCD background cross sections for the (S)LHC and VLHC, including
the signal K-factor, at the level of cuts in Eq.~(\ref{eq:cuts1}), adding
Eq.~(\ref{eq:cuts2}), and finally with all efficiencies and
misidentification probabilities applied, for both the conservative
(``hi'', $P_{j\to\gamma}=1/1600$) and optimistic (``lo'',
$P_{j\to\gamma}=1/2500$) assumptions. The QCD background normalization
uncertainty is rather large at LO, and unfortunately none of these
processes is known at NLO.  To estimate the effect of a possible
NLO increase of the background rates, we scale each of the background
cross sections by a factor~1.3.  Note that we are not making
any statement about unknown higher order corrections.  Instead, we
attempt to be conservative and show that our results do not critically
depend on the background normalization. 
\begin{table}[t]
\caption[]{\label{tab:xsec.l}
Expected cross sections [fb] (first three rows) for the $m_H=120$~GeV
$HH\to b\bar{b}\gamma\gamma$ signal and QCD backgrounds, including
the signal K-factors, at the (S)LHC. The background cross sections are
scaled by a factor 1.3, as explained in the text. The QCD backgrounds
cannot be calculated 
without cuts due to soft and collinear singularities.  Each of the
next four pairs of rows shows the cross sections including all
detector efficiencies and fake tag rejection probabilities as
described in the text, and the number of events expected, for each
machine and background analysis.  We assume an integrated luminosity
of 600~fb$^{-1}$ (6000~fb$^{-1}$) for the LHC (SLHC).  The $Hjj$,
$Hb\bar{b}$, $H\gamma\gamma$ and $Hj\gamma$ backgrounds are
discussed in the text and therefore not shown.}
\vspace{2mm}
\begin{tabular}{|c|c|c|c|c|c|c|c|c|c|c|c|}
analysis stage & $HH$
& $b\bar{b}\gamma\gamma$ & $c\bar{c}\gamma\gamma$
& $b\bar{b}\gamma j$ & $c\bar{c}\gamma j$ & $jj\gamma\gamma$
& $b\bar{b}jj$ & $c\bar{c}jj$ & $\gamma jjj$ & $jjjj$ & $\sum ({\rm bkg})$\\
\hline
before cuts & 0.15 & - & - & - & - & - & - & - & - & - & - \\
+ Eq.~(\ref{eq:cuts1})
& 0.043 & 0.056  & 0.42   & 65   & 250  & 11   & 2.5$\times 10^4$ & 2.5$\times 10^4$ & 7700 & 5$\times 10^6$ & 5$\times 10^6$ \\ 
+ Eq.~(\ref{eq:cuts2})
& 0.035 & 0.0060 & 0.0215 & 8.28 & 17.0 & 0.84 &  4520 &  4520 &  364 & 4$\times 10^5$ & 4$\times 10^5$ \\ 
\hline
\hline
$\times$ $\epsilon\cdot P^{hi}_{LHC}$ &
0.0106 & 0.0029 & 0.0020 & 0.0031 & 0.0013 & 0.0077 & 0.0013 & 0.0003 & 0.0030 & 0.0022 & 0.0233 \\
$N_{LHC\phantom{S}}(hi)$ & 6 &  2 &  1 &  2 &  1 &  5 &  1 &  0 &  2 &  1 & 14 \\
\hline
$\times$ $\epsilon\cdot P^{lo}_{LHC}$ &
0.0106 & 0.0029 & 0.0020 & 0.0020 & 0.0008 & 0.0077 & 0.0005 & 0.0001 & 0.0017 & 0.0009 & 0.0186 \\
$N_{LHC\phantom{S}}(lo)$ & 6 &  2 &  1 &  1 &  0 &  5 &  0 &  0 &  1 &  1 & 11 \\
\hline
\hline
$\times$ $\epsilon\cdot P^{hi}_{SLHC}$ &
0.0035 & 0.0010 & 0.0001 & 0.0010 & 0.0001 & 0.0010 & 0.0004 & 0.0000 & 0.0003 & 0.0003 & 0.0042 \\
$N_{SLHC}(hi)$ & 21 &  6 &  0 &  6 &  0 &  6 &  3 &  0 &  2 &  2 & 25 \\
\hline
$\times$ $\epsilon\cdot P^{lo}_{SLHC}$ &
0.0035 & 0.0010 & 0.0001 & 0.0007 & 0.0000 & 0.0010 & 0.0002 & 0.0000 & 0.0002 & 0.0001 & 0.0033 \\
$N_{SLHC}(lo)$ & 21 &  6 &  0 &  4 &  0 &  6 &  1 &  0 &  1 &  1 & 20 \\
\end{tabular}
\end{table}
\begin{table}[t]
\caption[]{\label{tab:xsec.v}
Expected cross sections [fb] (first three rows) for the $m_H=120$~GeV
$HH\to b\bar{b}\gamma\gamma$ signal and QCD backgrounds, including
the signal K-factors, at the VLHC.  The background cross sections are
scaled by a factor 1.3, as explained in the text. The QCD backgrounds
cannot be calculated 
without cuts due to soft and collinear singularities.  Each of the next
pairs of rows shows the cross sections including all detector
efficiencies and fake tag rejection probabilities as described in the
text, and the number of events expected for an integrated luminosity of
600~fb$^{-1}$, for each of the two background analyses.  We show results
for both single and double $b$-tagging.  The $Hjj$, $Hb\bar{b}$,
$H\gamma\gamma$ and $Hj\gamma$ backgrounds are discussed in the text and
therefore not shown.} 
\vspace{2mm}
\begin{tabular}{|c|c|c|c|c|c|c|c|c|c|c|c|}
analysis stage & $HH$
& $b\bar{b}\gamma\gamma$ & $c\bar{c}\gamma\gamma$
& $b\bar{b}\gamma j$ & $c\bar{c}\gamma j$ & $jj\gamma\gamma$
& $b\bar{b}jj$ & $c\bar{c}jj$ & $\gamma jjj$ & $jjjj$ & $\sum ({\rm bkg})$\\
\hline
before cuts & 15.9 & - & - & - & - & - & - & - & - & - & - \\
+ Eq.~(\ref{eq:cuts1})
& 3.12 & 1.8  & 23   & 3600 & 14000 & 280  & 1.6$\times 10^6$ & 1.6$\times 10^6$ & 2.3$\times 10^5$ & 2.7$\times 10^8$ & 2.7$\times 10^8$ \\ 
+ Eq.~(\ref{eq:cuts2})
& 2.70 & 0.14 & 1.23 &  417 &  1020 & 25.0 & 4.2$\times 10^5$ & 4.2$\times 10^5$ &  13300           & 3.0$\times 10^7$ & 3.0$\times 10^7$ \\ 
\hline
\multicolumn{12}{|c|}{1 $b$-tag}\\
\hline
$\times$ $\epsilon\cdot P^{hi}_{VLHC}$ &
0.810 & 0.067 & 0.116 & 0.156 & 0.075 & 0.228 & 0.122 & 0.024 & 0.095 & 0.164 & 1.048 \\
$N(hi)$ & 486 &  40 &  70 &  94 &  45 & 137 &  73 &  14 &  57 &  98 & 629 \\
\hline
$\times$ $\epsilon\cdot P^{lo}_{VLHC}$ &
0.810 & 0.067 & 0.116 & 0.100 & 0.048 & 0.228 & 0.050 & 0.010 & 0.061 & 0.067 & 0.747 \\
$N(lo)$ & 486 &  40 &  70 &  60 &  29 & 137 &  30 &   6 &  36 &  40 & 448 \\
\hline
\multicolumn{12}{|c|}{2 $b$-tags}\\
\hline
$\times$ $\epsilon\cdot P^{hi}_{VLHC}$ &
0.270 & 0.022 & 0.005 & 0.052 & 0.003 & 0.001 & 0.041 & 0.001 & 0.000 & 0.001 & 0.126 \\
$N(hi)$ & 162 &  13 &  3 &  31 &  2  & 0 &  25 &  1  &  0 &  1 & 76 \\
\hline
$\times$ $\epsilon\cdot P^{lo}_{VLHC}$ &
0.270 & 0.022 & 0.005 & 0.033 & 0.002 & 0.001 & 0.017 & 0.000 & 0.000 & 0.000 & 0.080 \\
$N(lo)$ & 162 &  13 &  3 &  20 &  1  & 0 &  10 &   0 &  0  &   0 & 47 \\
\end{tabular}
\end{table}

Before final state identification, the ${\cal O}(\alpha_s^4)$ $jjjj$
background dominates over all others by two orders of magnitude.  The
angular cuts of Eq.~(\ref{eq:cuts2}) do improve the signal to
background ratio by an order of magnitude, but it is the cumulative
effect of large rejection factors for misidentifying light jets as
photons or $b$-jets that brings the QCD backgrounds down to a
manageable level.

The single Higgs-resonance backgrounds are for the most part
negligible, so we do not include them in Tables~\ref{tab:xsec.l}
and~\ref{tab:xsec.v}.  The $Hjj$ cross section
is approximately a factor $6-20$ smaller than the signal~\cite{carlo}; the
$Hb\bar{b}$ cross section is a factor $20-60$ smaller.  Although no
calculations of $H\gamma\gamma$ and $Hj\gamma$ production exist yet,
one expects that these backgrounds are also negligible.  All
subsequent numerical results include the $Hjj$ background, whereas we
neglect the $Hb\bar{b}$, $H\gamma\gamma$ and $Hj\gamma$ backgrounds.\medskip

Summing all background cross sections we find that $S/B\sim 1/1$ is
possible at the SLHC, and we anticipate a still respectable $S/B\sim
1/2$ at the LHC.  At the VLHC, with one tagged $b$-quark, we obtain a
signal to background ratio of about $1/1$, while a double $b$-tag
yields to $S/B\approx 2 - 3.5$.  Of course, even small changes in
expected fake-$b$ rejection factors could change how the analysis
would be optimized.  Our results are meant only to highlight the
potential capability of such a search.

Our estimates also reveal that the range of fake photon rejection
probabilities is not so significant.  The largest background in most
cases is $jj\gamma\gamma$, where the photons are real but one or two
$b$-tags are falsely identified --- at the SLHC the double
$b$-tag requirement brings this background to the same level as the
real $b\bar{b}\gamma\gamma$ component.  The irreducible
$b\bar{b}\gamma\gamma$ background in all cases constitutes only a
small fraction of the total background.

As shown in Table~\ref{tab:xsec.v}, requiring two $b$-tags instead of
one at the VLHC reduces the overall background by a factor~8 --~9, but
the signal by only a factor~3.  As a result, both cases yield similar
sensitivity bounds for the Higgs self-coupling $\lambda$.  However, we
note that the higher event rate with one $b$-tag will provide better
control of experimental systematic uncertainties, so this may be the
preferred strategy.\medskip

In addition to the backgrounds considered so far,
$b\bar{b}\gamma\gamma$ events (or their fakes) may also be produced in
double parton scattering (DPS), or from multiple interactions occurring
from separate $pp$ collisions in the same bunch crossing at
high-luminosity running.  In principle, one can identify multiple
interactions by a total visible energy measurement or by tracing some
final particle tracks back to distinct event vertices, but this may
not always be possible in practice.  For example, for
$b\bar{b}\gamma\gamma$ events where the photon and $b\bar{b}$ pairs
occur in different interactions, the latter method relies solely on
tracks of particles associated with the hadronic activity accompanying
the photon pair.  If these particles are soft, the two vertices may
not be clearly resolvable.

To estimate the cross sections from DPS and multiple interactions, we
use the approximation outlined in Ref.~\cite{BCHP}. In both cases, the
dominant contribution arises from multi-jet production where several
jets are misidentified as $b$-quarks or photons.  After applying the
cuts listed in Eqs.~(\ref{eq:cuts1}) and~(\ref{eq:cuts2}), the DPS and
multiple interaction backgrounds are still several times larger than
the signal.  However, to discriminate them from regular single
interaction events, one can exploit the independence and pairwise
momentum balance of the two scatterings in DPS or multiple interaction
events, similar to the strategy employed in the DPS analysis carried
out by the CDF collaboration~\cite{dps}.  Rejecting events where two
sets of transverse momenta independently add up to a value close to
zero will obviously strongly suppress the DPS and multiple interaction
background.  The signal, on the other hand, is only minimally affected
by such a cut.  Requiring that events which pass the cuts listed in
Eqs.~(\ref{eq:cuts1}) and~(\ref{eq:cuts2}) do not satisfy either
\begin{equation}
|{\mathbf\vec{p}}_T(b)+{\mathbf\vec{p}}_T(\gamma_1)|<20~{\rm GeV}
\qquad {\rm and} \qquad 
|{\mathbf\vec{p}}_T(\bar{b})+{\mathbf\vec{p}}_T(\gamma_2)|<20~{\rm GeV}
\end{equation}
or
\begin{equation}
|{\mathbf\vec{p}}_T(\bar{b})+{\mathbf\vec{p}}_T(\gamma_1)|<20~{\rm GeV}
\qquad {\rm and} \qquad 
|{\mathbf\vec{p}}_T(b)+{\mathbf\vec{p}}_T(\gamma_2)|<20~{\rm GeV}
\end{equation}
totally eliminates the DPS and multiple scattering backgrounds (within
the limits of our ability to simulate detector effects), but reduces
the signal cross section by about $7\%$.  This has essentially no
influence on the Higgs self-coupling sensitivity bounds.\medskip

\begin{figure}[t!]
\begin{center}
\includegraphics[width=12cm]{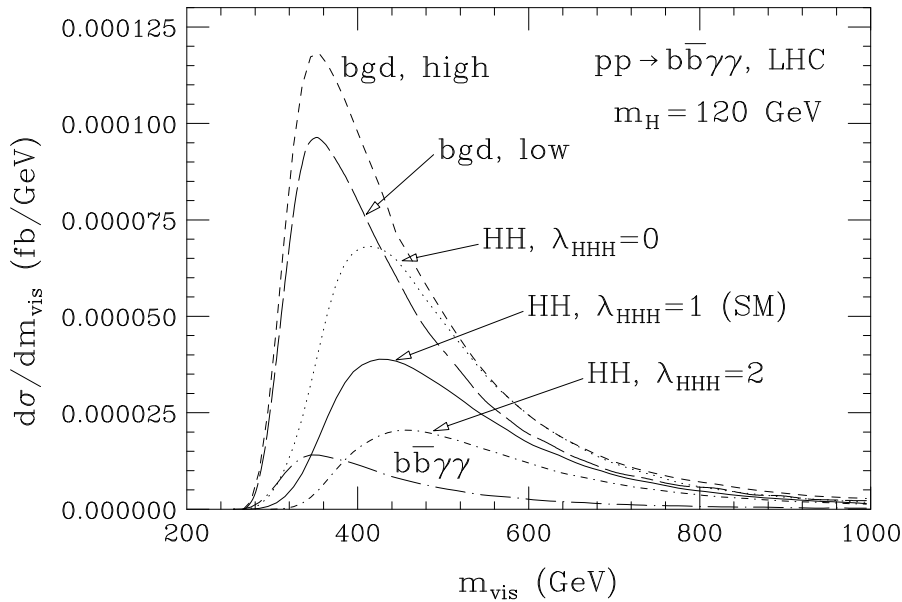}
\vspace*{2mm}
\caption[]{\label{fig:mvis-LHC}
  The visible invariant mass distribution, $m_{\rm vis}$, in $pp\to
  b\bar{b}\gamma\gamma$, after all kinematic cuts
  (Eqs.~(\protect\ref{eq:cuts1}) and~(\protect\ref{eq:cuts2})), for
  the conservative (short dashed) and optimistic (long dashed) QCD
  backgrounds and a SM signal of $m_H=120$~GeV (solid) at the LHC.
  The dotted and short dash-dotted lines show the signal cross section
  for $\lambda_{HHH}=\lambda/\lambda_{SM}=0$ and 2, respectively.  To
  illustrate how the reducible backgrounds dominate the analysis, we
  also show the irreducible QCD $b\bar{b}\gamma\gamma$ background by
  itself (long dash-dotted).  We include the NLO K-factor for the
  signal and a factor 1.3 for the QCD backgrounds.}  \vspace{-7mm}
\end{center}
\end{figure}

\begin{figure}[t!]
\begin{center}
\includegraphics[width=11cm,height=13cm]{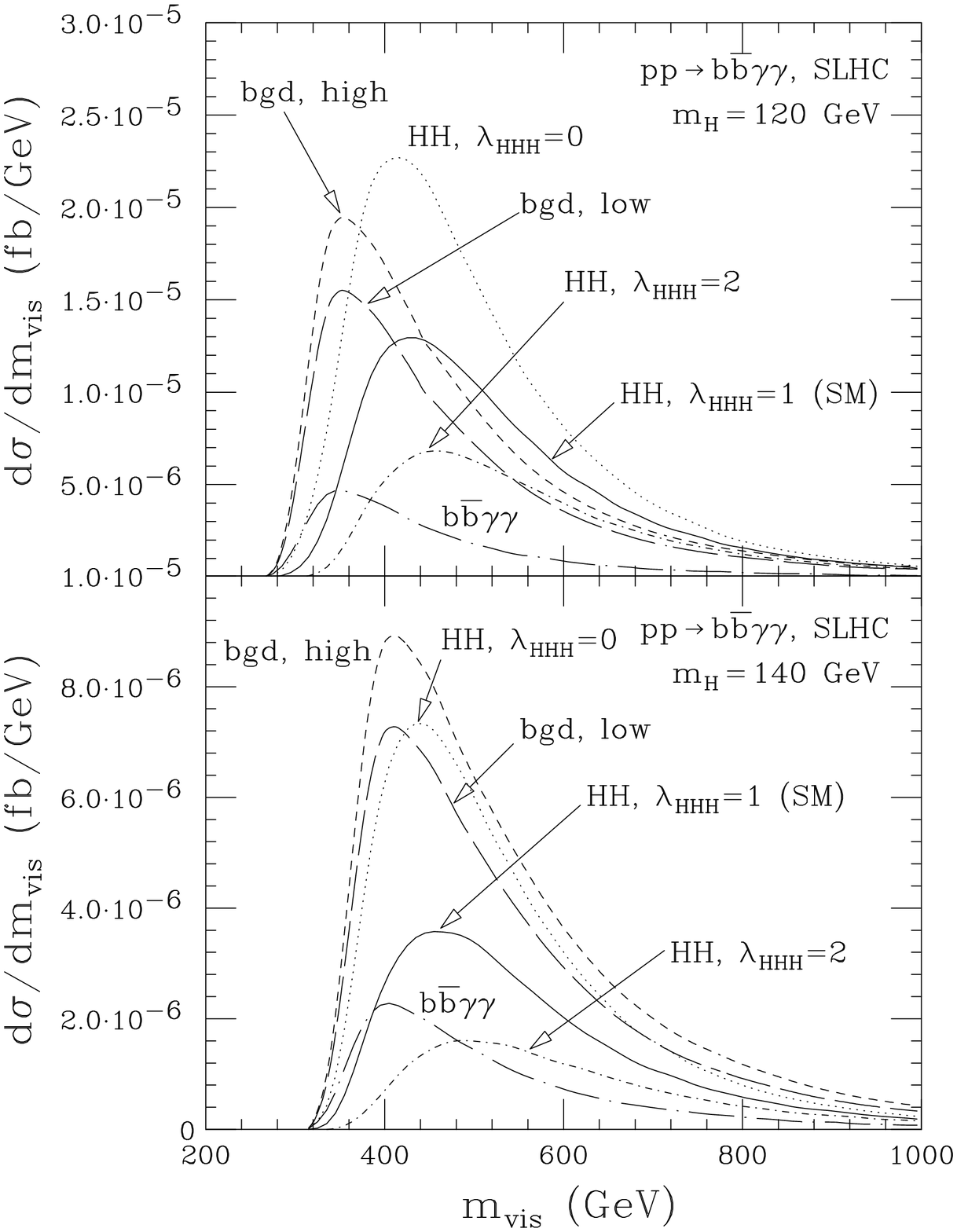}
\vspace*{2mm}
\caption[]{\label{fig:mvis-SLHC}
  The visible invariant mass distribution, $m_{\rm vis}$, in $pp\to
  b\bar{b}\gamma\gamma$, after all kinematic cuts
  (Eqs.~(\protect\ref{eq:cuts1}) and~(\protect\ref{eq:cuts2})), for
  the conservative (short dashed) and optimistic (long dashed) QCD
  backgrounds and SM signals of $m_H=120$ (upper) and 140~GeV (lower)
  at the SLHC.  The dotted and short dash-dotted lines show the signal
  cross section for $\lambda_{HHH}=\lambda/\lambda_{SM}=0$ and 2,
  respectively.  To illustrate how the reducible backgrounds dominate
  the analysis, we also show the irreducible QCD
  $b\bar{b}\gamma\gamma$ background by itself (long dash-dotted).  We
  include the NLO K-factor for the signal and a factor 1.3 for the QCD
  backgrounds.}  \vspace{-7mm}
\end{center}
\end{figure}

\begin{figure}[t!]
\begin{center}
\includegraphics[width=11cm,height=13cm]{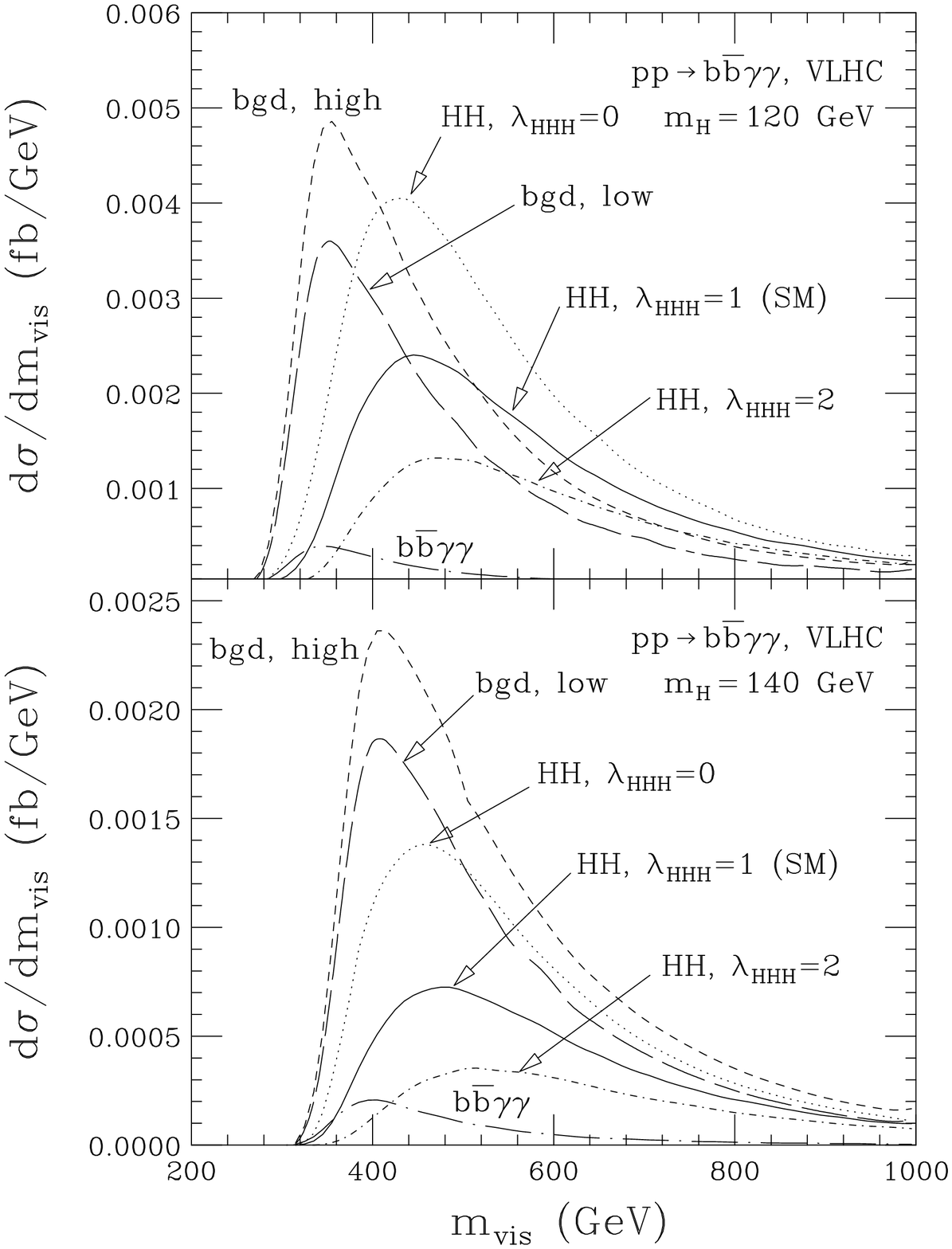}
\vspace*{2mm}
\caption[]{\label{fig:mvis-VLHC}
  The visible invariant mass distribution, $m_{\rm vis}$, in $pp\to
  b\bar{b}\gamma\gamma$, after all kinematic cuts
  (Eqs.~(\protect\ref{eq:cuts1}) and~(\protect\ref{eq:cuts2})), for
  the conservative (short dashed) and optimistic (long dashed) QCD
  backgrounds and SM signals of $m_H=120$ (upper) and 140~GeV (lower)
  at a VLHC.  The dotted and short dash-dotted lines show the signal
  cross section for $\lambda_{HHH}=\lambda/\lambda_{SM}=0$ and 2,
  respectively.  To illustrate how the reducible backgrounds dominate
  the analysis, we also show the irreducible QCD
  $b\bar{b}\gamma\gamma$ background by itself (long dash-dotted).  We
  include the NLO K-factor for the signal and a factor 1.3 for the QCD
  backgrounds.}  \vspace{-7mm}
\end{center}
\end{figure}

Extracting the Higgs boson self-coupling follows the same path as for
the $4W$ final state used for larger Higgs masses~\cite{BPR}.  To
discriminate between signal and background, we use the visible
invariant mass, $m_{\rm vis}$, which for this final state is the invariant
mass of the Higgs boson pair, corrected for energy loss of the
$b$-jets.  We show this in Fig.~\ref{fig:mvis-LHC} for $m_H=120$~GeV
at the LHC, and in Figs.~\ref{fig:mvis-SLHC} and~\ref{fig:mvis-VLHC}
for $m_H=120$~GeV and $m_H=140$~GeV at the SLHC and VLHC.
We do not show the $m_H=140$~GeV case for the LHC, since we expect
only about two signal events for an integrated luminosity of
600~fb$^{-1}$.  Figs.~\ref{fig:mvis-LHC} --~\ref{fig:mvis-VLHC} show
that the background distribution peaks close to the threshold, whereas
the signal distribution reaches its maximum at a somewhat higher
value.  This is due to the destructive interference between the
triangle and box diagrams contributing to $gg\to HH$.  It is
responsible for an increase in the signal cross section and a
shift in the $m_{\rm vis}$ peak position towards
lower values, if we assume $\lambda<\lambda_{SM}$,
and vice versa.  The shape of the visible invariant mass
distribution thus helps to discriminate signal and background and to
probe the Higgs self-coupling, $\lambda$.  Increasing $m_H$ from
120~GeV to 140~GeV reduces the signal (background) cross section by
about a factor~3 (2).\medskip

To derive quantitative sensitivity bounds on $\lambda$ we perform a
$\chi^2$ test of the $m_{\rm vis}$ distribution, similar to that
described in Ref.~\cite{BPR}.  Except for the Higgs self-coupling, we
assume the SM to be valid.  As in all previous analyses, we multiply
the LO differential cross sections of the QCD background processes by
a factor~1.3.  As mentioned before, this is not a guess of the higher
order corrections, which must either be computed, or the rates
measured sufficiently precisely.  However, this way we ensure that our
results do not critically depend on the absolute normalization of the
background rates, while of course they will depend on the uncertainty
associated with the determination of the background rate: we allow for
a normalization uncertainty of $10\%$ for the SM signal plus
background rate.  We express limits on the deviation of the Higgs
self-coupling from the SM value in terms of $\Delta\lambda_{HHH}$,
where
\begin{equation}
\Delta\lambda_{HHH}=\lambda_{HHH}-1={\lambda\over\lambda_{SM}}-1 \; .
\end{equation}
\begin{table}
\caption{Expected Higgs self-coupling $68.3\%$ CL ($1\sigma$)
sensitivity limits, expressed as
$\Delta\lambda_{HHH}={\lambda\over\lambda_{SM}}-1$, for the various
hadron collider options and background analyses presented in the text.
There are not enough events at the LHC for $m_H=140$~GeV to perform a
measurement of $\lambda$.  The LHC and VLHC analyses employ a
single $b$-tag strategy, while the high-luminosity conditions at the
SLHC force a double $b$-tag requirement.}
\label{tab:sum}
\vspace{2mm}
\begin{tabular}{ccccccc}
& \multicolumn{3}{c}{$m_H=120$~GeV} & \multicolumn{3}{c}{$m_H=140$~GeV} \\[1mm]
machine & ``hi'' & ``lo'' & bkg. sub. & ``hi'' & ``lo'' & bkg. sub. \\[0.5mm]
\tableline\\[-2mm]
LHC, 600~fb$^{-1}$    & $\matrix{ +1.9  \crcr\noalign{\vskip -4pt} -1.1  }$
                      & $\matrix{ +1.6  \crcr\noalign{\vskip -4pt} -1.1  }$
                      & $\matrix{ +0.94 \crcr\noalign{\vskip -4pt} -0.74 }$
                      & $\matrix{  -    \crcr\noalign{\vskip -4pt}  -    }$
                      & $\matrix{  -    \crcr\noalign{\vskip -4pt}  -    }$
                      & $\matrix{  -    \crcr\noalign{\vskip -4pt}  -    }$ \\[4mm]
SLHC, 6000~fb$^{-1}$  & $\matrix{ +0.82 \crcr\noalign{\vskip -4pt} -0.66 }$
                      & $\matrix{ +0.74 \crcr\noalign{\vskip -4pt} -0.62 }$
                      & $\matrix{ +0.52 \crcr\noalign{\vskip -4pt} -0.46 }$
                      & $\matrix{ +1.7  \crcr\noalign{\vskip -4pt} -0.9  }$
                      & $\matrix{ +1.4  \crcr\noalign{\vskip -4pt} -0.8  }$
                      & $\matrix{ +0.76 \crcr\noalign{\vskip -4pt} -0.58 }$ \\[4mm]
VLHC, 600~fb$^{-1}$   & $\matrix{ +0.44 \crcr\noalign{\vskip -4pt} -0.42 }$
                      & $\matrix{ +0.42 \crcr\noalign{\vskip -4pt} -0.40 }$
                      & $\matrix{ +0.32 \crcr\noalign{\vskip -4pt} -0.30 }$
                      & $\matrix{ +0.82 \crcr\noalign{\vskip -4pt} -0.62 }$
                      & $\matrix{ +0.66 \crcr\noalign{\vskip -4pt} -0.54 }$
                      & $\matrix{ +0.38 \crcr\noalign{\vskip -4pt} -0.34 }$ \\[4mm]
VLHC, 1200~fb$^{-1}$  & $\matrix{ +0.32 \crcr\noalign{\vskip -4pt} -0.30 }$
                      & $\matrix{ +0.30 \crcr\noalign{\vskip -4pt} -0.28 }$
                      & $\matrix{ +0.26 \crcr\noalign{\vskip -4pt} -0.22 }$
                      & $\matrix{ +0.76 \crcr\noalign{\vskip -4pt} -0.58 }$
                      & $\matrix{ +0.62 \crcr\noalign{\vskip -4pt} -0.50 }$
                      & $\matrix{ +0.36 \crcr\noalign{\vskip -4pt} -0.32 }$ \\[2mm]
\end{tabular}
\end{table}
We summarize our results in Table~\ref{tab:sum}.  The bounds obtained
using the conservative background estimate (labeled ``hi'') are
$10-20\%$ less stringent than those found using the more optimistic
scenario (labeled ``lo'').  At the SLHC, for $m_H=120$~GeV, a
vanishing Higgs self-coupling can be ruled out at the $90\%$~CL.
Limits for $m_H=140$~GeV are a factor~1.2 --~2 weaker than those for
$m_H=120$~GeV.

It may be possible to subtract large parts of the reducible
backgrounds which do not involve charm quarks using the following
technique.  Due to the their large cross sections (see
Tables~\ref{tab:xsec.l} and~\ref{tab:xsec.v}), one can fairly
accurately determine the $m_{\rm vis}$ distributions of the individual
processes, $Hjj$, $b\bar{b}\gamma j$, $b\bar{b}jj$, $jj\gamma\gamma$,
$\gamma jjj$ and $jjjj$ production, imposing the same cuts as in the
$HH\to b\bar{b}\gamma\gamma$ analysis (Eqs.~(\ref{eq:cuts1})
and~(\ref{eq:cuts2})).  If the photon--jet and light jet--$b$
misidentification probabilities are independently measured in
other processes such as prompt photon~\cite{wgam} and $W+$~jets
production, one can simply subtract these backgrounds.  For the
background processes involving charm quarks, on the other hand, this
procedure will be more difficult to realize, since the smaller charm
quark mass and the shorter charm lifetime result in a charm quark
tagging efficiency much lower than that for $b$-quarks.  The columns
labeled ``bgd. sub.'' list the limits achievable if the non-charm
reducible contributions to the background were subtracted with $100\%$
efficiency, but none of the charm quark backgrounds could be reduced.
Our results show that reducing the background beyond what can be
achieved with kinematic cuts may considerably improve the bounds on
$\lambda_{HHH}$ at the LHC and SLHC, where the $HH\to
b\bar{b}\gamma\gamma$ process is statistics limited.  The bounds
achievable at the SLHC (VLHC) by analyzing $b\bar{b}\gamma\gamma$
production are a factor~2.5 --~6 (2 --~3) more stringent than those
from the $b\bar b\tau^+\tau^-$ channel~\cite{Baur:2003gp}.

\smallskip

Due to the small number of events, the LHC and SLHC sensitivity limits
depend significantly on the SM cross section normalization
uncertainty.  For example, for a normalization uncertainty of $30\%$
on the SM signal plus background rate,
the achievable bounds on $\lambda_{HHH}$ are almost a factor~2 weaker
than those obtained for a normalization uncertainty of $10\%$. 
This SM cross section normalization
uncertainty depends critically on knowledge of
the QCD corrections to the signal and the ability to determine the
background normalization.  The NLO QCD corrections to $gg\to HH$ are
currently known only in the infinite top quark mass
limit~\cite{hh_nlo}.  To ensure the $10\%$ required precision on 
differential cross sections we would need the NLO rates for finite top 
quark masses, as well as the NNLO corrections in the heavy top quark 
mass limit.
For the background normalization one can rely on either
calculations of the QCD corrections or data.  As mentioned before,
none of these NLO background calculations are available.  Since there
are many processes contributing to the background, and most of them
involve hundreds of Feynman diagrams already at tree level, NLO
calculations appear feasible only if automated one-loop QCD tools
become available in the next few years.  In the absence of such NLO
results, one may be able to fix the background normalization instead
by relaxing the $b\bar{b}$ and $\gamma\gamma$ invariant mass cuts of
Eq.~(\ref{eq:cuts1}) and/or the cuts of Eq.~(\ref{eq:cuts2}) and
extrapolating from regions in $m_{b\bar{b}}$, $m_{\gamma\gamma}$,
$\Delta R(\gamma, b)_{min}$ and $\Delta R(\gamma,\gamma)$ where the
background dominates, back into the analysis region.  This technique
should make it possible to determine the background normalization to
about $10\%$ at the LHC and SLHC, and to about $2\%$ at the VLHC.
Both methods rely on Monte Carlo simulation to correctly predict the
$m_{\rm vis}$ distribution shape.

\smallskip

The bounds listed in Table~\ref{tab:sum} should be compared with those
achievable at $e^+e^-$ linear colliders.  A linear collider with
$\sqrt{s}=500$~GeV and an integrated luminosity of 1~ab$^{-1}$ can
determine $\lambda$ with a precision of about $20\%$ in $e^+e^-\to
ZHH$ for $m_H=120$~GeV~\cite{LC_HH4}.  For $m_H>120$~GeV, the $H\to
b\bar{b}$ branching ratio and the $e^+e^-\to ZHH$ cross section both
fall off quickly.  Since the background
cross section decreases only slightly, $S/B$, and thus the bounds on
$\lambda$ obtainable from $e^+e^-\to ZHH$, worsen rapidly with
increasing values of $m_H$.  By $m_H=140$~GeV they are at only the
$50\%$ level~\cite{Baur:2003gp}.  From Table~\ref{tab:sum} it is clear
that the LHC will be able to provide only a first rough measurement of
the Higgs self-coupling for $m_H=120$~GeV.  A luminosity-upgraded LHC
will be able to make a more precise measurement.  However, the
sensitivity bounds on $\lambda$ obtained from $b\bar{b}\gamma\gamma$
production for $m_H=120$~GeV ($m_H=140$~GeV) will be a factor~2 --~4
(1.2 --~3) weaker than those achievable at a linear collider.  In
contrast, the sensitivity at a VLHC will approach this level of
precision.  It should be noted that if the SM cross section
normalization uncertainty could be reduced to a few percent, a VLHC
could reach precision similar to that foreseen for CLIC~\cite{LC_HH3}
($e^+e^-$ collisions at 3~TeV center-of-mass energy).

\subsection{The $b\bar{b}\mu^+\mu^-$ decay channel}

The $b\bar{b}\mu^+\mu^-$ signal calculation proceeds as in the
$b\bar{b}\gamma\gamma$ case.  The basic kinematic acceptance cuts for
events at the LHC and VLHC are:
\begin{eqnarray}
\label{eq:mucuts}
\nonumber &
p_T(b) > 45~{\rm GeV} \; , \qquad
|\eta(b)| < 2.5 \; , \qquad
\Delta R(b,b) > 0.4 \; , \\
\nonumber &
m_H-20~{\rm GeV} \, < \, m_{b\bar{b}} \, < \, m_H+20~{\rm GeV} \; , \\
\nonumber &
p_T(\mu) > 15~{\rm GeV} \; , \qquad
|\eta(\mu)| < 2.4 \; , \qquad
\Delta R(\mu,\mu) > 0.4 \; , \\
\nonumber &
m_H-5~{\rm GeV} \, < \, m_{\mu\mu} \, < \, m_H+5~{\rm GeV} \; , \\
&
\Delta R(b,\mu) > 0.4 \; ,
\end{eqnarray}
where again the muon invariant mass window is chosen to accept $79\%$
of the $H\to\mu^+\mu^-$ decay after detector effects.  The signal
cross section at the LHC (VLHC) for $m_H=120$~GeV before taking into
account any efficiencies is 2.4~ab (0.21~fb), approximately one order
of magnitude smaller than the $b\bar{b}\gamma\gamma$ channel.  For
larger Higgs boson masses the ratio is even smaller, due to the
$H\to\mu^+\mu^-$ branching ratio, which decreases much more rapidly
with $m_H$ than that for $H\to\gamma\gamma$ (see Fig.~\ref{fig:BRs}).
Once efficiencies are taken into account, we expect less than one
signal event at the LHC.  The SLHC would see 2 --~3 signal events for
$m_H=120$~GeV if one assumes that both $b$-quarks are tagged, too few
for a meaningful coupling extraction.  At a VLHC there would be about
60 signal events for an integrated luminosity of 600~fb$^{-1}$, single
$b$-tag requirement, and the same value of $m_H$.  We therefore
concentrate on the VLHC in the following, and require only one
$b$-tag.\smallskip

A potential advantage of the $b\bar{b}\mu^+\mu^-$ final state is the
smaller number of processes contributing to the background.  The main
contributions to the background originate from QCD
$b\bar{b}\mu^+\mu^-$, $c\bar{c}\mu^+\mu^-$ and $jj\mu^+\mu^-$
production, where the $\mu^+\mu^-$ pair originates from an off-shell
$Z$-boson or photon.  In the latter two processes, either a charm
quark or light jet is misidentified as a $b$-quark.  We calculate the
background processes at LO using {\sc MCFM}~\cite{mcfm} and find that
their sum is more than a factor~200 larger than the signal.  The
signal to background ratio improves by a factor~5 if we additionally
require
\begin{equation}
\Delta R(\mu,\mu)<2 \; ,
\end{equation}
whereas the signal cross section falls by only about $20\%$.  The
$Hjj$ background is negligible compared with $jj\mu^+\mu^-$.  The
final signal to background ratio of $S/B\approx 1/50$ contrasts starkly
with the $S/B\sim 1/1$ ratio the $b\bar{b}\gamma\gamma$ channel
enjoys.  If instead both $b$-jets are tagged, the signal to background
ratio improves by an additional factor~2.  However, the signal cross
section is reduced by a factor~3, which yields sensitivity bounds
for $\lambda_{HHH}$ which are somewhat weaker than those obtained
from single $b$-tag data.

Shrinking the $\mu^+\mu^-$ invariant mass window could also reduce the
background.  The value in Eq.~(\ref{eq:mucuts}) was chosen assuming
ATLAS detector muon momentum resolution~\cite{atlas_tdr}.  The CMS
detector~\cite{cms_tdr} likely can use a smaller window,
$|m_H-m_{\mu\mu}| < 3$~GeV, which would reduce the background by
approximately a factor~1.7.\smallskip

The small signal cross section combined with the very large background
make it essentially impossible to determine the Higgs boson
self-coupling in $pp\to b\bar{b}\mu^+\mu^-$.  We quantify this by
performing a $\chi^2$ test on the $m_{\rm vis}$ distribution, similar to
that described in Sec.~\ref{sec:gamgam}.  Since the signal cross
section is too small to be observable at the LHC and SLHC, we derive
bounds only for a VLHC.  As before, we include the effects of NLO QCD
corrections via multiplicative factors: $K=1.35$ for the signal~\cite{hh_nlo},
$K=0.81$ for $b\bar{b}\mu^+\mu^-$ and $c\bar{c}\mu^+\mu^-$ production,
and $K=0.91$ for the $jj\mu^+\mu^-$ background~\cite{mcfm}.  Allowing
for a normalization uncertainty of $10\%$ of the SM cross sections,
for $m_H=120$~GeV we find $1\sigma$ bounds of
\begin{equation}
-3.0 < \Delta\lambda_{HHH}< 4.2
\label{eq:mulim}
\end{equation}
at the VLHC for an integrated luminosity of 600~fb$^{-1}$. If the
$jj\mu^+\mu^-$ background can be subtracted as described in
Sec.~\ref{sec:gamgam}, the limits improve by about a factor~1.4.
Using the CMS dimuon mass window instead, the bound improves by about
a factor~1.3.  Nevertheless, this is about an order or magnitude
weaker than the limits from $HH\to b\bar{b}\gamma\gamma$.

\section{Supersymmetric Higgs Bosons}
\label{sec:mssm}

The MSSM requires two Higgs doublets, in contrast to one in the SM, to
give mass to the up--type and the down--type fermions and to avoid
anomalies induced by the supersymmetric fermionic partners of the
Higgs bosons.  This results in the presence of five physical Higgs
bosons: a charged pair $H^\pm$, two neutral scalars $h^0$ and $H^0$,
and a pseudoscalar $A^0$.  The two scalars are mixed mass eigenstates,
the lighter always having a mass $m_h\lesssim
135$~GeV~\cite{higgsmass}.  At leading order, the entire MSSM Higgs
sector is described by two parameters, usually taken to be the ratio
of the two Higgs doublets' vacuum expectation values, $\tan\beta$, and
the pseudoscalar Higgs mass, $m_A$.  In the region $m_A \gtrsim
150$~GeV, all heavy Higgs bosons $A,H,H^\pm$ have similar masses, much
larger than the light scalar Higgs mass. In this so--called decoupling
regime the light Higgs boson $h$ strongly resembles a SM Higgs boson
of the same mass.  It will be difficult to distinguish between the SM
and the MSSM Higgs sectors through measurements of its
properties~\cite{wbf_exp,Hcoup}.

Assuming bottom--tau mass unification, only two regions of $\tan\beta$
are allowed: either small values, $\tan\beta\lesssim 3$, or large
values, $\tan\beta\gtrsim 30$.  Direct searches for the heavy Higgs
bosons are particularly promising in the large $\tan\beta$ regime,
since in the decoupling limit the bottom Yukawa coupling to heavy
Higgses is $m_b\tan\beta$. As a result, $b$-quark initiated processes,
such as $b\bar{b}\to H$, may have cross sections enhanced by up to
three orders of magnitude over the corresponding SM rates for
sufficiently large values of $\tan\beta$. In contrast, for small
values of $\tan\beta$ these direct searches fail, because the dominant
Yukawa coupling becomes $m_t/\tan\beta \gg m_b \tan\beta$.\smallskip

At the LHC, associated production of two neutral MSSM Higgs bosons via
gluon fusion occurs for all six possible combinations~\cite{lhc_hh}.
In principle, these processes probe the various Higgs boson
self-couplings, $\lambda_{ijk}$. However, for large $\tan\beta$ the
continuum box diagrams are enhanced by the Yukawa coupling squared,
while the triangle loop diagram with an intermediate Higgs boson is
enhanced by only one power of the large Yukawa coupling: for large
$\tan\beta$ the resonance diagrams are suppressed by $1/\tan\beta$ as
compared to the continuum production diagrams. For $\tan\beta=50$ we
find that the effect of vanishing self couplings $\lambda_{ijk}\equiv
0$ is at maximum at the percentage level.\medskip

For $\tan\beta \gtrsim 30$ and $m_A\lesssim 150$~GeV, MSSM Higgs pair
production cross sections can be sizable, reaching values up to
100~fb, compared to a few tens of fb in the SM.  The largest cross
sections occur for two heavy states $AH,AA,HH$ and large values of
$\tan\beta$, due to the enhanced coupling of these states to
$b$-quarks.  In this regime the most promising final state is
$b\bar{b}\mu^+\mu^-$ since the ratio of the muon and the bottom Yukawa
couplings is preserved in the MSSM, but the branching ratio to photons
is highly suppressed, typically by several orders of magnitude
compared to the SM Higgs boson of equal mass.  Unfortunately, a main
background for this is MSSM $b\bar{b}H/A,H/A\to\mu^+\mu^-$
production~\cite{eduard}.  Whether the Higgs pair signal could be
extracted out of this would require a more detailed investigation
which we do not find likely to be fruitful.\medskip

\begin{figure}[t]
\begin{center}
\includegraphics[width=11cm]{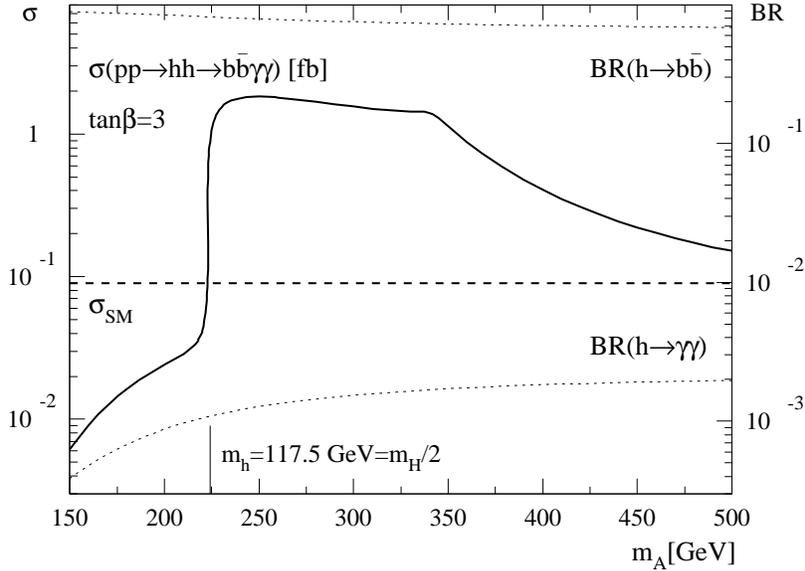}
\vspace{3mm}
\caption[]{\label{fig:sig-susy} Lowest order cross section and
  branching fractions for pair production of light MSSM scalar Higgs
  bosons, $pp\to hh$, with subsequent decay $hh\to
  b\bar{b}\gamma\gamma$, as a function of the pseudoscalar Higgs mass
  $m_A$.  We fix $\tan\beta=3$, set the squark mass parameters to
  1~TeV, and assume maximal mixing with
  $A_t=2.5$~TeV~\cite{feynhiggsfast}.  We do not take into account
  supersymmetric decay modes of the heavy Higgs boson
  $H$~\cite{hdecay}.  The light Higgs boson mass is above the LEP
  limit of $m_H>114.4$~GeV~\cite{lep} for $m_A>190$~GeV.  No cuts or
  detection efficiencies are included. The dashed horizontal line shows
  the lowest order SM $gg\to HH$ cross section for $m_H=120$~GeV.
}\vspace{-7mm}
\end{center}
\end{figure}

In the small $\tan\beta$ regime it is much more difficult to
distinguish the SM and the MSSM Higgs sectors.  None of the heavy
Higgs bosons will be directly observable at the LHC for
$\tan\beta\lesssim 20$, if we rely on the usual decays to fermions.
We find that, for small values of $\tan\beta$, $gg\to H\to hh$ offers
the best chance to detect the heavy scalar Higgs boson, $H$: for
$\tan\beta\lesssim 5$ the $H\to hh$ 
branching ratio is sizable~\cite{susy_4b}.  To take into account
off--shell effects we compute the full $pp\to hh$ production rate.  As
in the SM, we expect the 
$b\bar{b}\gamma\gamma$ final state to be most promising in the
decoupling regime, with increased rate due to the intermediate $H$
resonance.  We show the $h\to b\bar{b}$ and $h\to\gamma\gamma$
branching fractions and lowest order $gg\to hh\to
b\bar{b}\gamma\gamma$ cross section as a function of $m_A$ in
Fig.~\ref{fig:sig-susy}.  The light Higgs boson mass increases from
$m_h=108$~GeV for $m_A=150$~GeV to a plateau value of $m_h=122$~GeV in
the large $m_A$ limit.  A few structures in the cross section plot
require further explanation.  First, the heavy scalar Higgs mass
crosses the threshold $m_H>2m_h$ around $m_A\sim 225$~GeV, which
enhances the $hh$ cross section by almost a factor 100.  Second, the
kink at $m_A\sim m_H=350$~GeV represents the top threshold in the top
triangle loop. At the same time we see the onset of the $H\to
t\bar{t}$ decay channel, which for larger values of $m_A$ dominates
over $H\to hh$, so the cross section decreases rapidly.  Nevertheless,
the MSSM signal rate is still enhanced over the SM rate
$\sigma_{SM}(b\bar{b}\gamma\gamma)\approx 0.09$~fb for values of $m_A$
as large as 500~GeV.

\begin{figure}[t]
\begin{center}
\includegraphics[width=11cm]{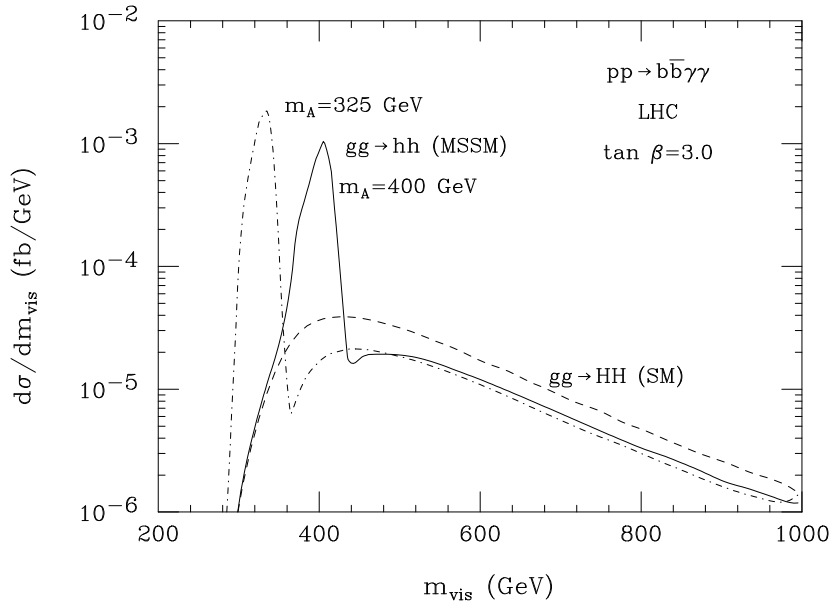} \vspace*{2mm}
\caption[]{\label{fig:mvis-susy} The visible invariant mass
  distribution, $m_{\rm vis}$, for MSSM light scalar Higgs pair
  production at the LHC, $pp\to hh\to b\bar{b}\gamma\gamma$, for
  $\tan\beta=3$. The light Higgs mass for $m_A=325$~GeV is 120.8~GeV
  and for $m_A=400$~GeV it is 122.2~GeV.  For comparison, we also show
  the distribution for SM Higgs pair production
  ($m_H=120$~GeV).}\vspace{-7mm}
\end{center}
\end{figure}

Unfortunately, the angular cuts of Eq.~(\ref{eq:cuts2}) which are
needed to suppress the background, together with the standard
$b\bar{b}\gamma\gamma$ identification cuts of Eq.~(\ref{eq:cuts1}),
force the differential cross section to vanish for $m_{\rm
vis}\lesssim 250$~GeV. Pair production of light supersymmetric Higgs
bosons will thus be unobservable for $m_A<280$~GeV.  When taking into
account detection efficiencies, we find that $hh$ production at the
LHC should be observable at the $5\sigma$ level for $320<m_A<375$~GeV
($310<m_A<425$~GeV) for an integrated luminosity of 300~fb$^{-1}$
(600~fb$^{-1}$) and $\tan\beta=3$.  The signal would be rather
spectacular: due to $s$-channel $H$ exchange, the differential cross
section peaks for $m_{\rm vis}\approx m_H$, as shown in
Fig~\ref{fig:mvis-susy}. Compared to the SM case the cross section is
enhanced by more than an order of magnitude in the resonance region,
where it depends on the $Hhh$ and $Hf\bar{f}$ couplings. Since MSSM
heavy scalar $H$ production with decay into fermions is unobservable
at the LHC in the small $\tan\beta$ region, this implies that $hh$
production can measure only a combination of $\lambda_{Hhh}$ and the
$Hf\bar{f}$ couplings, but not the individual couplings.

%

\section{Discussion and Conclusions}
\label{sec:conc}

After discovery of an elementary Higgs boson and
tests of its fermionic and gauge boson couplings, experimental
evidence that the shape of the Higgs potential has the form required
for electroweak symmetry breaking will complete the proof that fermion
and weak boson masses are generated by spontaneous symmetry breaking.
One must determine the Higgs self-coupling to probe the shape of the
Higgs potential.\smallskip

Only Higgs boson pair production at colliders can accomplish this.
Numerous studies~\cite{LC_HH1,LC_HH2,LC_HH3,LC_HH4} have established
that future $e^+e^-$ machines can measure $\lambda$ at the $20-50\%$
level for $m_H<140$~GeV.  Very recent studies~\cite{SLHC,BPR,blondel}
determined that the prospects at hadron colliders for $150<m_H<200$~GeV
are similarly positive, but that the $m_H<140$~GeV 
region would be very difficult to access~\cite{Baur:2003gp}.  We have
tried to rectify the situation in this paper by considering highly
efficient, lower-background rare decay modes: $b\bar{b}\gamma\gamma$
and $b\bar{b}\mu^+\mu^-$.  The latter suffers from very low rate and
considerable background from the Breit-Wigner tail of $b\bar{b}Z$
production, and does not appear to be useful.  This is not surprising
upon comparison to our $b\bar{b}\tau^+\tau^-$
study~\cite{Baur:2003gp}.\smallskip

However, the $b\bar{b}\gamma\gamma$ channel shows considerable
promise.  Imposing photon--photon and photon--$b$ separation cuts
could result in a signal to background ratio of
${\cal O}(1)$ or better.  Since the irreducible QCD
$b\bar{b}\gamma\gamma$ background is small compared to the reducible
background originating from light jets or charm quarks mistagged as
$b$-quarks, or from jets misidentified as photons, the signal to
background ratio depends on the particle misidentification
probabilities, and the required number of $b$-tags.

We find that the LHC, with an integrated luminosity of 600~fb$^{-1}$
or more, could make a very rough first measurement for $m_H=120$~GeV
(with $\sim 6$ signal events), but would not obtain useful limits for
$m_H=140$~GeV at all due to the lack of signal events.  It would
require a luminosity-upgraded run (SLHC, 6000~fb$^{-1}$) to rule out
$\lambda=0$ at the $90\%$ CL for $m_H=120$~GeV, and to make a
$50-80\%$ measurement at the $1\sigma$ level.  A 200~TeV VLHC, in
contrast, would make possible a $20-40\%$ measurement of $\lambda$,
competitive with future $e^+e^-$ collider capabilities.  We note,
however, that current understanding of hadron collider Higgs boson
phenomenology doesn't provide for the necessary precision knowledge of
Higgs branching ratios to complement this.  It is likely that an
$e^+e^-$ collider would still be required to fill this role.  Although
a luminosity-upgraded LHC cannot compete with a linear collider for
Higgs masses $m_H<140$~GeV, a Higgs self-coupling measurement at the
SLHC will still be interesting if realized before a linear collider
begins operation.\medskip

To fully exploit future hadron collider potential to measure the Higgs
self-coupling, we need an accurate prediction of the SM
$b\bar{b}\gamma\gamma$ rate.  It is mandatory that the residual
theoretical cross section uncertainty be reduced to the $10-15\%$
level for any $HH$ analysis to be meaningful.  We will need similar
precision on background rates probably from experiment by
extrapolating from background-dominated phase space regions to that of
the signal.\bigskip

Probably the most exciting result of this analysis is the MSSM case:
the heavy MSSM Higgs scalar can decay into two light Higgs bosons if
$\tan\beta\lesssim 5$.  This region of parameter space poses a serious
challenge to the LHC, because none of the usual heavy Higgs searches
will detect a hint of the two Higgs doublets required in the MSSM.
Resonant production of the heavy scalar Higgs in gluon fusion and its
subsequent decay into light Higgs bosons, which then decay to
$b\bar{b}\gamma\gamma$, has two effects on the cross section as
compared to the SM case: the total rate is enhanced by about an order of
magnitude 
and the $hh$ invariant mass peaks at the heavy Higgs mass. Even though
our analysis is not at all optimized for resonant MSSM production, we
find a $5\sigma$ discovery region for $\tan\beta=3$ and $310<m_A<
425$~GeV at the LHC. Even though the discovery reach of this channel
does not extend to much larger values of $\tan\beta$ it still ensures
the observation of one heavy Higgs boson in a region preferred by
bottom--tau unification, inaccessible by other MSSM Higgs searches.

\acknowledgements

We would like to thank K.~Bloom, M.~D\"uhrssen, F.~Maltoni,
B.~Mellado, A.~Nikitenko, J.~Parsons, D.~Wackeroth, D.~Zeppenfeld and
P.M.~Zerwas for useful discussions. We also thank C.~Oleari for
providing us with code to calculate the $Hjj$ background. One of us
(U.B.)  would like to thank the Fermilab Theory Group, where part of
this work was carried out, for its generous hospitality.  This
research was supported in part by the National Science Foundation
under grant No.~PHY-0139953.



\bibliographystyle{plain}

\end{document}